\documentclass{ifacconf}
\usepackage{graphicx}     
\usepackage{natbib}       
\usepackage{graphics} 
\usepackage{subcaption}
\usepackage{epsfig} 
\usepackage{mathptmx} 
\usepackage{times} 
\usepackage{amsmath} 
\usepackage{amssymb} 
\usepackage[dvipsnames]{xcolor}
\usepackage{mathtools}
\usepackage{tikz}
\usetikzlibrary{arrows,chains,matrix,positioning,scopes}
\usepackage{chemmacros}
\newcommand{\rvline}{\hspace*{-\arraycolsep}\vline\hspace*{-\arraycolsep}}

\begin{document}
\begin{frontmatter}

\title{Global and local synaptic regulation determine the stability of homeostatic plasticity \thanksref{footnoteinfo}}

\thanks[footnoteinfo]{S. Aljaberi is supported by Abu Dhabi National Oil Company (ADNOC). A. Bellotti is supported by the Gates Cambridge Trust and the NIH OxCam Scholars Program. T O'Leary is supported by ERC grant StG 716643 FLEXNEURO.  }

\author[First]{Saeed Aljaberi} 
\author[First]{Adriano Bellotti} 
\author[First]{Timothy O'Leary}
\author[First]{Fulvio Forni}

\address[First]{Department of Engineering, University of Cambridge, UK (e-mail: sa798/ ab2424/timothy.oleary/f.forni@eng.cam.ac.uk). }

\begin{abstract}                
 Neurons regulate the distribution of signaling components across an extended tree-like cellular structure using both local and global feedback control. This is hypothesized to allow homeostatic control of the electrical activity of a neuron and at the same time enable normalization of distribution of inputs received from other cells. The performance and robustness of these mechanisms are poorly understood, and are subject to nonlinearities, making their analysis difficult. Firstly, we formally show that \emph{global} homeostasis of electrical activity and \emph{local} activity-dependent degradation can coexist under sufficient timescale separation. The interplay of the two feedback mechanisms is also analyzed through simulations, which reveal a bidirectional effect (stabilizing and destabilizing) of activity-dependent degradation on the overall neuron performance.
\end{abstract}


\end{frontmatter}

\section{Introduction}
Neurons are electrically excitable cells that receive input from potentially thousands of other cells via elaborate tree-like dendrites. The ion channels and receptors that receive and process signals in dendrites have finite lifespans of days or hours, and need to be continually replenished \citep{marder2006variability}. Neurons therefore maintain electrical activity by regulating the synthesis of these components. Furthermore, the relative strengths of inputs to neurons continually adapt as a mean of storing information. It is widely hypothesized that neurons implement an approximate weight normalization of their inputs to preserve information while avoiding signal saturation, a process known as synaptic scaling \citep{turrigiano2008self}.

The adaptive nature of neurons and the finite lifespan of their signaling components suggests that feedback control is essential to maintain the function of a neural circuit. Much of the basic physiology of this feedback has been experimentally characterized, consisting of a biochemical sensor that reads out average electrical activity by detecting calcium influx \citep{o2014cell,o2010homeostasis}. This activity readout is used as a feedback signal to control the rate of synthesis of ion channel and receptor proteins. However, the size and complexity of neuronal dendritic trees indicates that such regulation is far from trivial. Proteins and protein precursors (mRNAs) need to be actively synthesized and transported over potentially large distances. Many (but not all) components are synthesized at the cell body, or soma, and then actively transported along a microtubule network that traverses the dendritic tree \citep{Burute_2019,bressloff2009cable}. This can result in significant delays between the synthesis of a component and its arrival at a site where it is needed \citep{williams2016dendritic}.

Experimental data suggest two broad classes of activity-dependent feedback mechanisms in neurons: global and local feedback. Global feedback regulates the synthesis of material at the cell body, necessarily including all mRNA synthesis. Local feedback regulates the synthesis and delivery of proteins in the vicinity of the site of use throughout the dendritic tree \citep{fernandez2014meet,glock2017mrna,fonkeu2019mrna}. The role of both mechanisms in maintaining neuronal function is the subject of intense experimental research and debate, and is believed to vary substantially across biological contexts, including animal species, brain area and neuron type. Nonetheless, these mechanisms serve two broad goals:
\begin{enumerate}
	\item[(i)] the \emph{global} task of maintaining average electrical activity at an (approximate) set point;
	\item[(ii)] the \emph{local} task of supporting heterogeneous distribution of receptors and ion channels across the dendritic tree.
\end{enumerate}
There is no system-theoretic analysis of the contribution of global and local feedback mechanisms to both goals. Therefore, there is a gap in our understanding of how the division of labor between these mechanisms constrains the robustness and flexibility of neuronal regulation.

Firstly, to fulfill goal (i), we formulate a controller that depends on the average readout of ion channels in the network, and show that there is a fundamental constraint on the maximum allowable feedback gain (Theorem \ref{thm1}). Secondly, to fulfill goal (ii), we propose a distributed set of controllers that depends on the local ion channel concentration. We show that the two feedback mechanisms can coexist to support a stable behavior (Theorem \ref{thm:glocal}). Finally, through simulations, we illustrate how the interaction of global and local action exhibits both stabilizing and destabilizing effects on the closed loop system performance.

The manuscript is organized as follows. In Section \ref{model} we develop a closed-loop dendritic trafficking model. Section \ref{fin} characterizes the stability of global and local feedback control. Section \ref{syn} shows how interactions between global and local feedback affects system behavior, focusing on imposed changes in the spatial distribution of receptors in the dendrites. Conclusions follow. Proofs are in the last section of the paper. 

\section{Neural transport and homeostasis}
\label{model}	
\subsection{Neuronal transport}
\begin{figure}[htbp]
	\centering
	\includegraphics[width=2.9in]{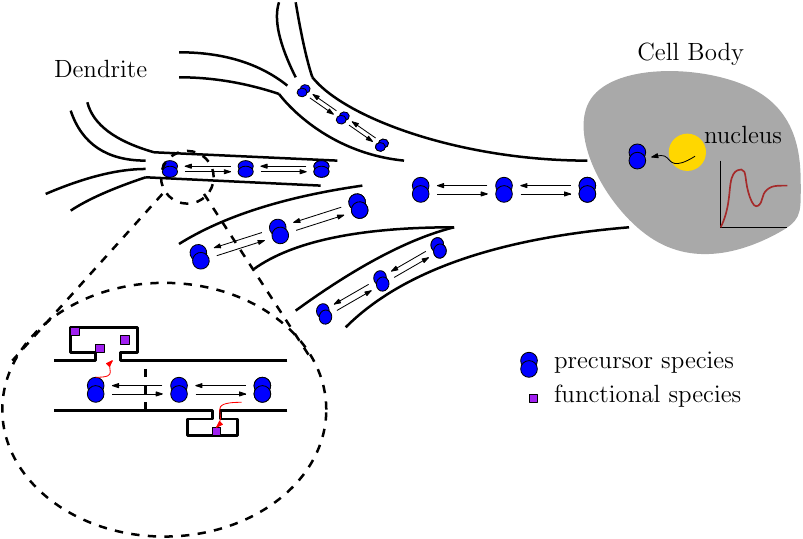}
	\caption{Schematic of a neuron showing its structure and sites where the processes of interest occur, from the cell body to the extremities of the dendritic tree.} 
	\label{fig:schematic}
\end{figure}

A simplified sketch of a neuron is shown in Figure 1. We model the neuron as an interconnection of $n$ compartments.
The first represents the soma (or cell body), while the remaining compartments refer to sections of the dendritic tree. 
Using $m_i$ to denote the concentration of material in compartment $i$, the variation in time of ${m}_{i}$ is described by 
\begin{align}
\dot{m}_{i} = & -(\omega^{m}_{i} + \sum_{j=1,i\ne j}^{n}v_{ij}){m}_{i} + \sum_{j=1,j\ne i}^{n}v_{ji}{m}_{j} + b_{i}u \nonumber
\end{align}
where $v_{ij} \geq 0 $ is the trafficking rate or speed of material moving from compartment $i$ to compartment $j$, 
and $\omega^{m}_{i} > 0$ is the degeneration rate of ${m}_{i}$. We assume $v_{ij}=0$ if and only if $i$ is not connected to $j$. The input $u$ represents $m_{i}$ synthesis or production. We assume that material production occurs primarily in the cell body, where the required machinery exists, including nucleic acids and ribosomes. Moreover, we are interested in an isolated neuron and therefore ignore exogenous sources of $m_{i}$. This leads to a nonzero production only in first compartment, i.e. $b_{i} = 1$  when $i = 1$. Overall, the dynamics are represented by
\begin{align}
\label{EQ:m_matrix}
\dot{m} = & (L - \Omega_{m})m + Bu
\end{align}
where $m\in \mathbb{R}^{n}$, $L \in \mathbb{R}^{n\times n}$, and $\Omega_{m}=\text{diag}\{\omega^{m}_{i}\}$. 
The off-diagonal elements of $L$ satisfy $l_{ij}=v_{ji}$.
The diagonal elements of $L$ satisfy $l_{ii} = - \sum_{j=1,j\ne i}v_{ij}$. 

\eqref{EQ:m_matrix} is a drift-diffusion system modeling active intracellular transport, performed by motor proteins
(\cite{williams2016dendritic,aljaberi2019qualitative}). To allow the material to reach every point in the dendrite, we assume that there is alway a path from the first compartment to any other compartment.
Finally, note that if $l_{ij}\ne 0$ then $l_{ji} \ne 0$.

For the type of phenomena we are interested in studying, the key molecule defining the material concentration $m_i$ in each
compartment $i$ is likely to undergo a series of biochemical/biophysical reactions or structural changes. 
Examples of such changes could be transcription-translation, phosphorylation, or detachment of cargo from a motor-cargo complex. 
This motivates the introduction of a second species, which will be referred to as ${g}_{i}$. 
The main difference between the two species is that ${m}_{i}$ undergoes transport while ${g}_{i}$ doesn't. 
In this modeling framework, ${m}_{i}$ are the precursor species, while ${g}_{i}$ are the functional species. The dynamics of $g_{i}$ are
described by the differential equation $\dot{g}_{i} = s_{i}{m}_{i} - \omega^{g}_{i}{g}_{i}$ 
where $s_{i}$ is the transformation factor from ${m}_{i}$ to ${g}_{i}$ 
and $\omega^{g}_{i}$ is the degradation rate of ${g}_{i}$. In matrix form,
\begin{align}
\label{g_matrix}
\dot{g} = & S{m} - \Omega_{g}g
\end{align}
where $g \in \mathbb{R}^{n}$, $S=\text{diag}\{s_{i}\}$, $\Omega_{g} = \text{diag}\{\omega^{g}_{i}\}$, and $S, \Omega_{g} \in \mathbb{R}^{n\times n}$. 

\subsection{Electrical activity and homeostasis}
In our model we assume that the material ${m}_{i}$ is an mRNA-type molecule and 
that $g_i$ is the concentration of a ion-channel type molecule in the neuron's membrane, 
whose role is to shape the electrical features of the neuron.
We model the neuron as a leaky-integrator, considering the standard single compartment membrane equation
$C\dot{V} {}={}    g_{leak}(E_{leak} - V) + G(E_{g} - V)$, where
$V$ is the membrane potential, $C$ is membrane capacitance, $g_{leak}$ is a fixed, leak conductance, $E_{leak}$ and $E_g$  are equilibrium potentials, and 
\begin{equation}
\label{eq:G}
G = c^{T}g \ , \qquad c \in \mathbb{R}^n
\end{equation}
represents the weighted average of the ion-channel molecule concentrations.
By using a single compartment membrane equation, we assume that the neuron is equipotential ($V$ is independent of compartment index). We further assume timescale separation between the fast voltage fluctuations and the 
synthesis / trafficking of ${m}$ / $g$ dynamics. This allow us to approximate the membrane potential to its quasi-steady state
\begin{equation}
\label{EQ:V}
V := V_{\text{ss}} =  \frac{ G E_{g} + g_{leak} E_{leak}}{g_{leak} + G}.
\end{equation}
\eqref{EQ:V} shows how the weighted concentration of ion-channels $G$ affects the electrical
activity of the neuron. Existing models posit that biochemical pathways regulate
the synthesis of ion-channels to preserve a steady average electrical activity in the neuron. Specifically, we assume that this average electrical activity is mediated by calcium concentration
\begin{equation}
\label{EQ:Ca}
[Ca^{2+}] =  \frac{\alpha}{ 1 + \exp{(-V/\beta)}}
\end{equation}
where $\alpha,\beta > 0$ describe voltage sensitivity and calcium buffering of calcium channels 
  (\cite{o2013correlations}). 
 The regulation mechanism penalizes the deviation of calcium concentration $[Ca^{2+}]$ 
from an effective set-point, $[Ca^{2+}]_{\text{target}}$. 
The form of the control action that transforms the error signal
\begin{equation}
\label{EQ:e_G}
e_{G} = [Ca^{2+}]_{\text{target}} - [Ca^{2+}]
\end{equation}
into the production rate $u$ in \eqref{EQ:m_matrix} is the subject of ongoing research. 
Here we consider a simple leaky-integral control
\begin{equation}
\label{EQ:leaky-integrator}
\dot{u} = k_{G}e_{G} - \gamma_{G} u 
\end{equation}	
where $\gamma_{G} > 0$ sets a small degradation rate, and $k_{G}$ is the feedback (integral) gain. 

\begin{figure}[htbp]
	\centering
	\includegraphics[width=3.2in]{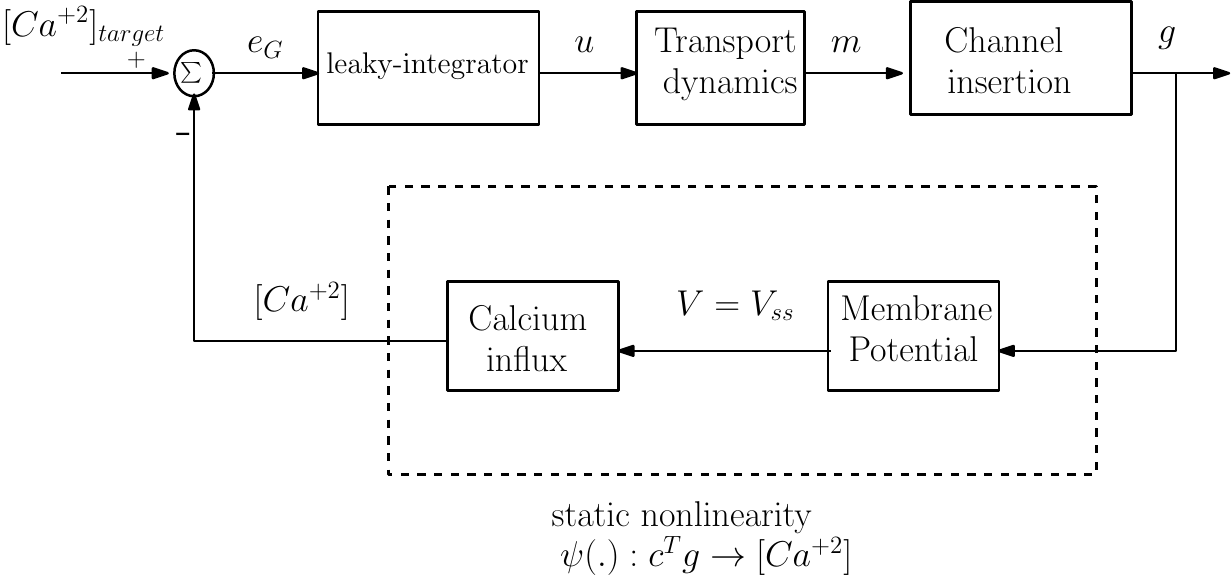}
	\caption{Block diagram of the closed-loop dendritic trafficking with global controller.}
	\label{fig:block_G}
\end{figure}

The main elements of the model  are summarized in the block diagram 
in Figure \ref{fig:block_G}. Note that \eqref{EQ:V}-\eqref{EQ:Ca} are lumped in the nonlinear map 
$\psi(.):c^{T}g\rightarrow [Ca^{2+}]$. 

The following assumption guarantees that $[Ca^{2+}] = [Ca^{2+}]_{\text{target}}$ 
is a feasible objective. 

\textit{Standing Assumption}.
The parameters of \eqref{EQ:V}-\eqref{EQ:Ca} satisfy 
$$E_{g} + \beta\ln \frac{\alpha}{[Ca^{2+}]_{\text{target}}} \ne 0 \mbox{ and } g_{leak} \ne 0.$$

Furthermore, closed-loop stability is achieved for every configuration of 
parameters and for every morphology
of the neuron, provided that the feedback gain is sufficiently small. 
\begin{thm}
\label{thm1}
	For any selection of system parameters, there exists a feedback gain
	$\bar{k}_G>0$ such that  the equilibrium of the 
	closed-loop  model \eqref{EQ:m_matrix}-\eqref{EQ:leaky-integrator} is globally exponentially
	stable for any $0 < k_G \leq \bar{k}_G$.
\end{thm}

Theorem \ref{thm1} holds even if we assume that the $g_{i}$'s are undergoing transport, i.e. if equation \eqref{g_matrix} was
$
\dot{g} =  S{m} - (L_{g} + \Omega_{g})g
$
where $L_{g}$ now captures the trafficking among the $g$ species.

\section{Ion channel density regulation}
\label{fin}

\subsection{Limitation of the homeostatic controller}
\label{sec:global_limitations}
\begin{figure}[htbp]
	\begin{center}
	\includegraphics[width=0.9\columnwidth]{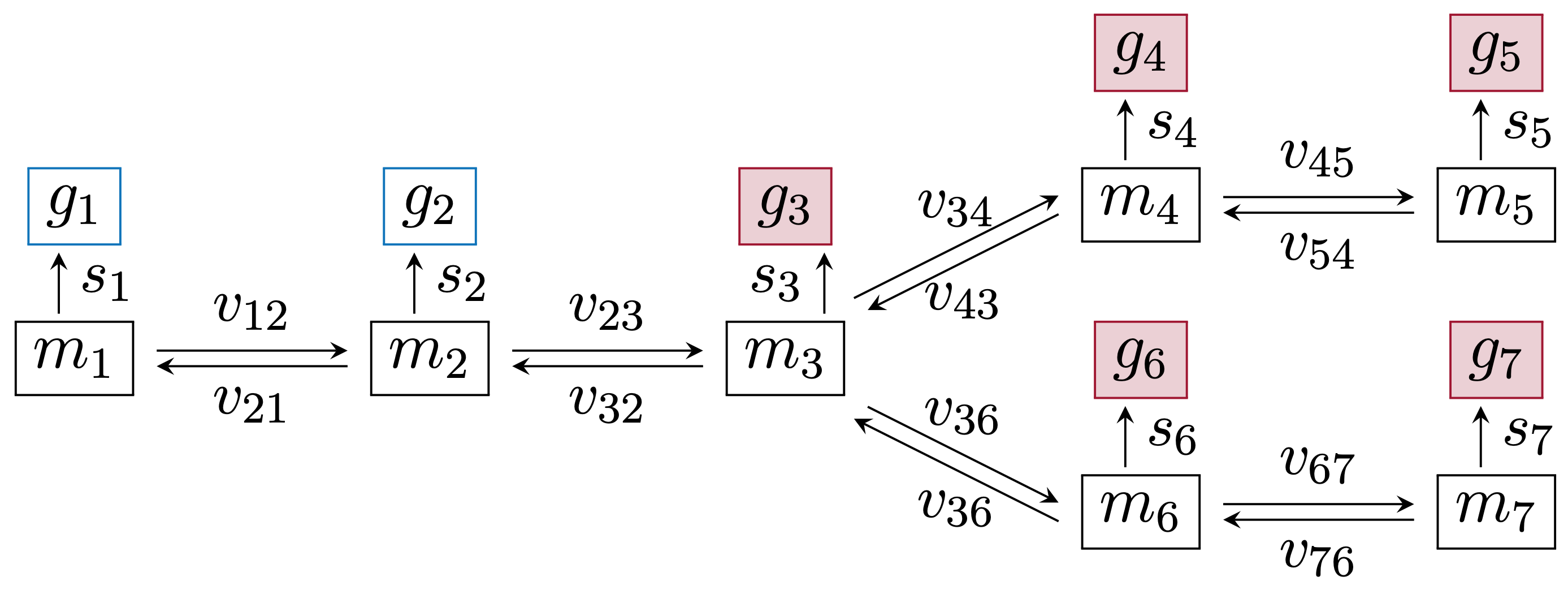} \vspace{4mm}
	
			{\footnotesize\begin{tabular}{ |c|c|c|c| } 
			\hline
			$v_{ij}=1$ if $i<j$ & $v_{ij}=0.5$ if $i>j$ & $\omega^{m}_{i} = \omega^{g}_{i} = 0.1$& $g_{leak}=0.25$ \\ 
			$E_{leak}=-50$ & $E_{g}=20$ & $\alpha=1$&  $[Ca^{2+}]_{\text{target}} = 0.5$\\
			$\beta=1$ & $\gamma_{G} = 0.0001$ & $k_{G}=$ varies &  $s_{i}=1$  \\
			\hline
		\end{tabular}}
			\caption{Neuron topology  and parameters.} 	\label{fig:ex_g}
		\end{center}

\end{figure}

\begin{figure}[htbp]
	\centering
	\begin{subfigure}[htbp]{\columnwidth}
		\centering
		\includegraphics[width=0.45\linewidth]{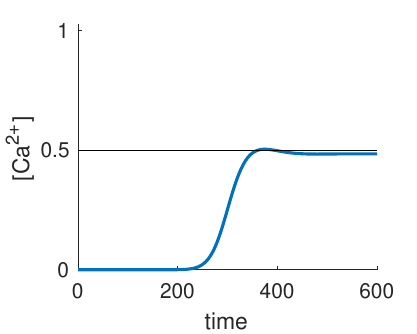}%
		\hfill
		\includegraphics[width=0.45\linewidth]{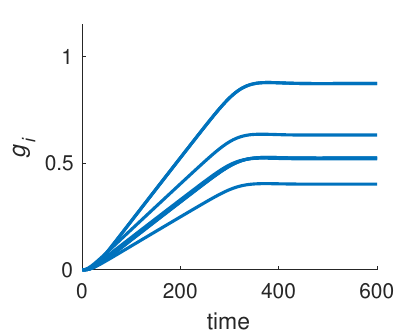}
		\caption{$k_{G} = 0.003$}
	\end{subfigure}
	\vskip\baselineskip
	\begin{subfigure}[htbp]{\columnwidth}
		\centering
		\includegraphics[width=0.45\linewidth]{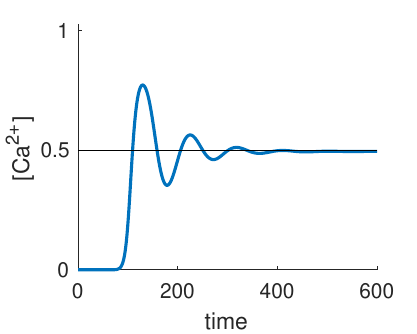}%
		\hfill
		\includegraphics[width=0.45\linewidth]{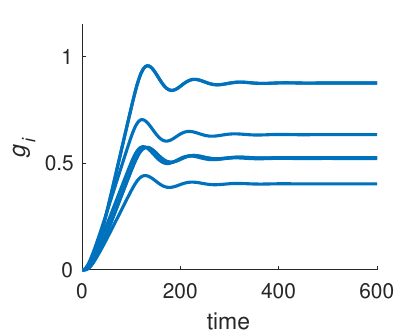}
		\caption{$k_{G} = 0.01$}
	\end{subfigure}
	\begin{subfigure}[htbp]{\columnwidth}
		\centering
		\includegraphics[width=0.45\linewidth]{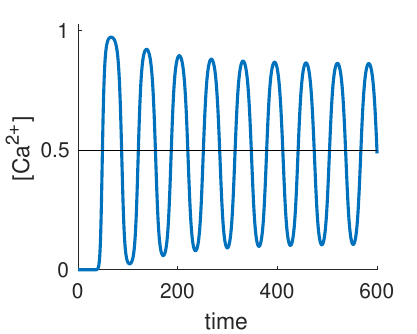}%
		\hfill
		\includegraphics[width=0.45\linewidth]{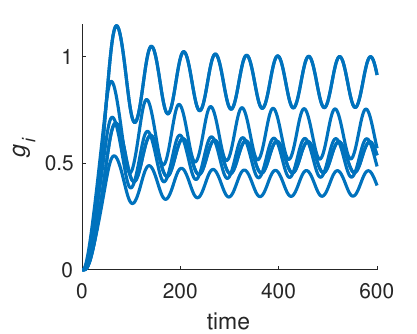}
		\caption{$k_{G} = 0.03$}
	\end{subfigure}
	\caption{Response of the closed loop \eqref{EQ:m_matrix}-\eqref{EQ:leaky-integrator} in Figure \ref{fig:ex_g}.}
	\label{fig:ex_G_sim}
\end{figure}

The closed-loop dendritic trafficking model achieves stable
regulation, with the presence of degradation leading to imperfect tracking. However, the overall architecture suffers from several limitations.
This is illustrated via simulation based on topology and parameters in Figure \ref{fig:ex_g}.
For increasing values of $k_G$, the simulations 
 in Figure \ref{fig:ex_G_sim} reveal that the system is 
well-behaved and achieves regulation for small values of $k_G$.
Convergence improves for larger values of the feedback gain.
Eventually, however, high values of feedback gain 
lead to instability (oscillations). 


For any generic selection of parameters, if the control gain is sufficiently small, Theorem \ref{thm1} guarantees that
the homeostatic controller guarantees stable average electrical activity of the neuron, a process known as synaptic scaling \citep{costa2009translational,turrigiano2008self}. This means that the controller necessarily tolerates changes to the distribution of receptors and ion channels that are imposed by other processes, such as long-term potentiation and depression, while gradually normalizing activity to a target level. However, it cannot shape the overall distribution of ion channels in the dendritic tree. 
The variability in ion-channel distribution is another important physiological characteristic in dendrites that reflects cognitive tasks, such as storing information. Hence, the closed-loop model \eqref{EQ:m_matrix}-\eqref{EQ:leaky-integrator} only fulfills goal (i), but not goal (ii).

\subsection{Distributed adaptation}
Taking inspiration from \citep{fernandez2014meet,glock2017mrna,fonkeu2019mrna}, 
we propose a distributed adaptation 
mechanism to enable fine tuning of ion-channel concentrations. 
This is implemented through adaptation of the 
degradation rates $\omega^{m}_{i}$ in feedback from the local 
concentrations ${g}_{i}$. The adaptation is to 
penalize the mismatch from a desired target while maintaining 
(physiological) positive degradation rates.
The mechanisms is defined by the following basic model
\begin{equation}
\label{w_compact}
\begin{split}
i \in \mathbb{I}: \quad 
& \epsilon \dot{\omega}^{m}_{i}   
= -\gamma_{L}\omega^{m}_{i} + k_L(g_i-\bar{g}_i) + \phi_{\ell b}(\omega^{m}_i,\bar\omega)  \\
i \not\in  \mathbb{I}: \quad 
&  \omega^{m}_{i}   
= \omega^{m}_{i,const} 
\end{split}
\end{equation}
where $ \mathbb{I}$ is an index set identifying the compartment with active adaptation,
$\gamma_{L} > 0$, $k_L>0$, $\omega^{m}_{i,const} \geq \bar\omega>0$, and $\varepsilon >0$ are generic parameters;
$\bar{g}_{i}$ is the desired ion-concentration set point, and
$\phi_{\ell b}(\cdot,\bar\omega)$ is a differentiable decreasing barrier function, with domain $(\bar\omega,\infty)$, 
whose role is to guarantee that the adaptation of $\omega^{m}_i$ never goes below the boundary $\bar{\omega}$.
$\phi_{\ell b}(\omega_i,\bar\omega)$ should not affect the dynamics away from the boundary $\bar{\omega}$,
as clarified below.

\textit{Standing Assumption.} 
For any given $0 <  \varepsilon_{\bar\omega} \ll 1$, we assume that 
the barrier function $\phi_{\ell b}(\cdot,\bar\omega)$ satisfies
$\phi_{\ell b}(\omega^{m}_i,\bar\omega) = 0$ and 
$\frac{d}{d \omega^{m}_i} \phi_{\ell b}(\omega^{m}_i,\bar\omega) = 0$ for all $\omega^{m}_i \geq \bar\omega + \varepsilon_{\bar\omega}$. 

The time constant 
$\epsilon $ is typically small, to reflect the fact that the 
adaptation of $\omega^{m}_{i}$ occurs at a faster timescale
than \eqref{EQ:m_matrix}-\eqref{EQ:leaky-integrator}.
Standard singular perturbation methods  lead 
to the following result (see e.g.  \cite[Theorem 11.4]{khalil2002nonlinear})

\begin{thm}
\label{thm:glocal}
For any given selection of system parameters, suppose that 
there exists intervals 
$0 < k_G \leq \bar{k}_G$ and  $0 < \varepsilon \leq \bar{\varepsilon}$
for which the equilibria $(m^*,g^*,u^*,\omega^*)$ of the closed-loop system 
\eqref{EQ:m_matrix}-\eqref{w_compact} satisfy 
$\omega_i^* \geq \bar\omega + \varepsilon_{\bar\omega}$.
Then, for $k_G$  and $\varepsilon$ sufficiently small, the 
closed loop equilibrium is exponentially stable.
\end{thm}

Theorem \ref{thm:glocal} makes clear that the combination of 
global homeostasis and distributed adaptation 
guarantees stability of the equilibrium of the system, which depends
on the local concentrations targets, $\bar{g}_i$, and of the 
calcium  target, $[Ca^{2+}]_{\text{target}}$.

\begin{figure}[htbp]
	\centering
	\begin{subfigure}[htbp]{\columnwidth}
		\centering
		\includegraphics[width=0.45\linewidth]{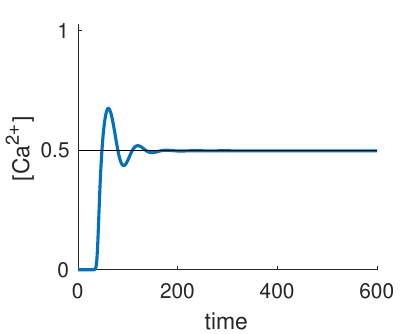}%
		\hfill
		\includegraphics[width=0.45\linewidth]{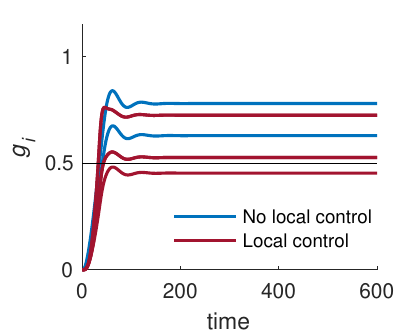}
		\caption{$k_{G} = 0.03$, $k_L = 1$, $\gamma_L = 1$.}
	\end{subfigure}
	\vskip\baselineskip
	\begin{subfigure}[htbp]{\columnwidth}
		\centering
		\includegraphics[width=0.45\linewidth]{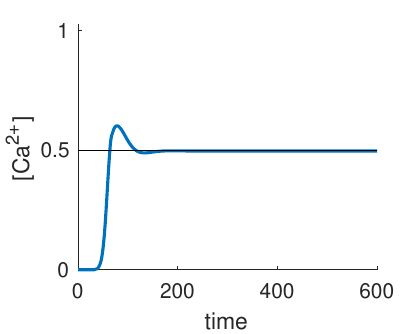}%
		\hfill
		\includegraphics[width=0.45\linewidth]{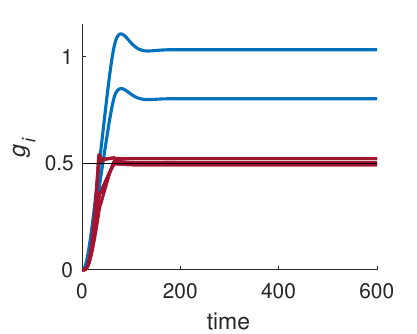}
		\caption{$k_{G} = 0.03$, $k_L = 20$, $\gamma_L = 1$.}
	\end{subfigure}
	\caption{The  closed loop \eqref{EQ:m_matrix}-\eqref{w_compact} for $\bar{g}_{i} = 0.5$ and $\epsilon = 0.1$.}
	\label{fig:ex_GI_sim}
\end{figure}

\begin{rem}
A more biologically plausible modeling choice is when \eqref{w_compact} depends on local calcium concentration, $[Ca^{+2}]_{i}$, which in turn depends on $g_{i}$  ($[Ca^{2+}]_{i} = f(g_{i})$). However, the behavior of $f(g_{i})$ is usually monotone, akin to that of the somatic intracellular calcium map $ [Ca^{2+}] = \psi(c^{T}g)$. In other words, one expects the dependence to involve a series of sequestration processes and signaling pathways in the form of a Hill equation or Michaelis Menten kinetics. Therefore, we expect the qualitative features of the closed loop system to remain unchanged, and a we adopt the dependence on $g_{i}$ as in \eqref{w_compact} for simplicity.
\end{rem}
As a matter of illustration, we revisit the simulations
in Section \ref{sec:global_limitations}, to show the effectiveness of the adaptation mechanisms
to fine tune the distribution of ion-channels. 
Adaptation is applied to compartments $3-7$, with  local set points $\bar{g}_{i} = 0.5$. 
Figure \ref{fig:ex_GI_sim} shows how local set points are recovered for increasing
values of the local gain $k_L$. Moreover, the local adaptation has a stabilizing effect on the
closed-loop dynamics, substantially reducing system oscillations for $k_G = 0.03$.

\begin{figure}[htbp]
		\centering
		\includegraphics[width=0.45\linewidth]{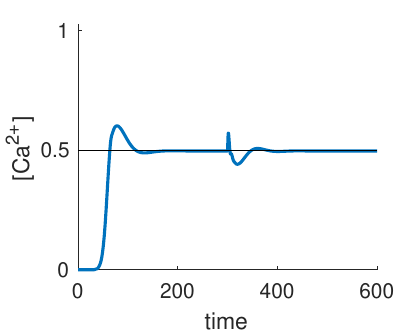}%
		\hfill
		\includegraphics[width=0.45\linewidth]{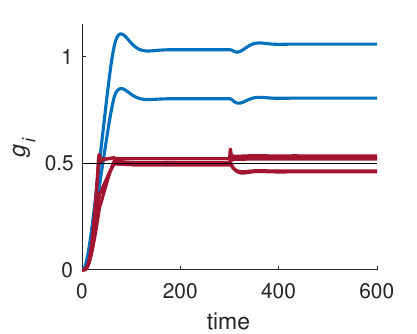}
	\caption{Perturbed response of the closed loop \eqref{EQ:m_matrix}-\eqref{w_compact} for $k_G = 0.03$,
	$k_L = 20$, and $ \gamma_L = 1$. At $t = 300$,  $s_{4}$  and  $s_{5}$ switch from $1$ to $2$. The system recovers.}
	\label{fig:s}
\end{figure}

The simulations in Figure \ref{fig:s} further illustrate the effectiveness of the local action to cope
with a perturbation on the rates $s_{4}$ and $s_{5}$ occurring at $t = 300$. 
After a brief transient, the desired set point is restored.

\section{Synaptic plasticity and competition}
\label{syn}
Connections between neurons can change in strength during learning, a process known as synaptic plasticity. For our purposes we view these changes as being introduced exogenously, resulting in a local increase or decrease in the concentration of receptors at the membrane, where communication with other cells can occur. We therefore distinguish between the concentrations of receptors that are being transported within the neuron ($m$) from the distribution of receptors at the the membrane ($g$). In order to understand how distributions in $g$ are affected by global and local feedback, we assume that the input from other cells has a constant (uniform) average value.

In this setting, the global controller \eqref{EQ:leaky-integrator} is responsible for synaptic homeostasis, which sets the average number of receptors in the system. On the other hand, the local action sets the local set point in response to changes, such as a learning event. In some cases, there might be a competition between the global and local objectives, where not all $g$ profiles are achievable. 

\begin{figure}
	\begin{subfigure}[htbp]{\columnwidth}
		\centering
		\includegraphics[width=0.45\linewidth]{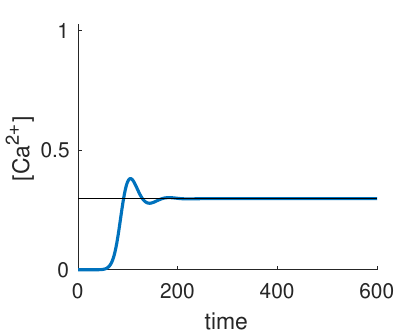}%
		\hfill
		\includegraphics[width=0.45\linewidth]{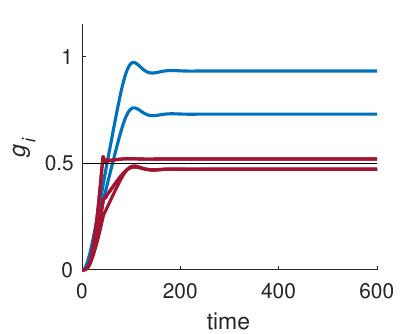}
		\caption{$[Ca^{2+}]_{\text{target}} = 0.3$}
		\label{fig:sp2}
	\end{subfigure}
	\begin{subfigure}[htbp]{\columnwidth}
		\centering
		\includegraphics[width=0.45\linewidth]{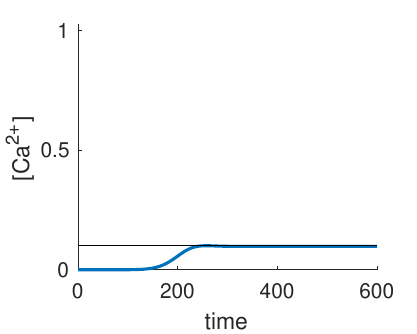}%
		\hfill
		\includegraphics[width=0.45\linewidth]{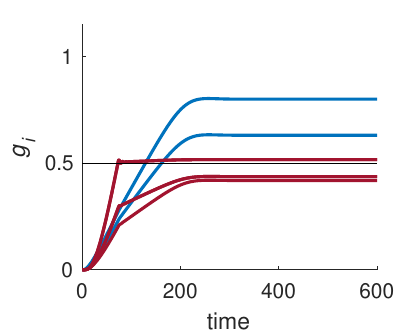}
		\caption{$[Ca^{2+}]_{\text{target}} = 0.1$}
		\label{fig:sp3}
	\end{subfigure} 
	\begin{subfigure}[htbp]{\columnwidth}
		\centering
		\includegraphics[width=0.45\linewidth]{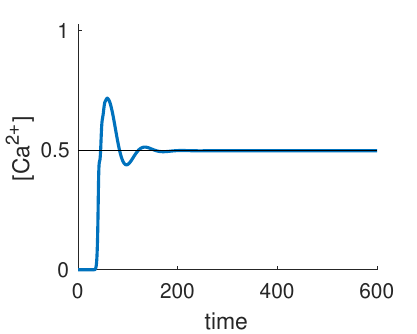}%
		\hfill
		\includegraphics[width=0.45\linewidth]{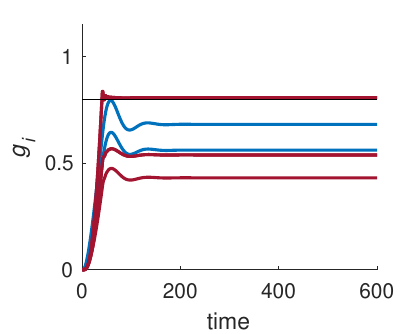}
		\caption{$\bar{g}_{i} = 0.8$}
		\label{fig:sp5}
	\end{subfigure}
	\begin{subfigure}[htbp]{\columnwidth}
		\centering
		\includegraphics[width=0.45\linewidth]{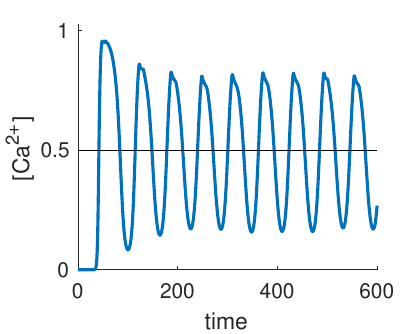}%
		\hfill
		\includegraphics[width=0.45\linewidth]{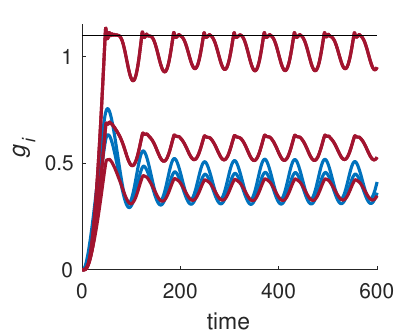}
		\caption{$\bar{g}_{i} = 1.1$}
		\label{fig:sp6}
	\end{subfigure} \vspace{-2mm}
	\caption{(a)-(c): decreasing global set point and (d)-(f): increasing local set point
	for 	$k_G = 0.03$, $k_L = 20$, $\gamma_L = 1$.}
	\label{fig:synaptic_plasticity} 
\end{figure}

Figures \ref{fig:sp2}-\ref{fig:sp3} show the system response as $[Ca^{2+}]_{\text{target}}$ is gradually decreased.
Decreasing $[Ca^{2+}]_{\text{target}}$ corresponds to decreasing the total amount of receptors in the system, hence making it difficult for the local action to succeed in achieving $\bar{g}_{i}$. When the local action fails to achieve the local set-point, it achieves the closest possible steady state given the limited total mass in the system.
The tug-of-war between the global controller and local action can ultimately lead to instability. This is illustrated by gradually increasing the local set point while maintaining a fixed Calcium target, as shown in Figures \ref{fig:sp5}-\ref{fig:sp6}.

\section{Conclusions}
\label{fut}
The paper presented a closed-loop model of dendritic trafficking, combining a homeostatic mechanism with a distributed adaptation to regulate both average electrical activity of the neuron and the distribution of ion channels across the dendritic tree. Using singular perturbation arguments, the paper provides stability guarantees on the system behavior. The features of the closed loop are also discussed, through simulations. 
In particular, we showed how the distributed adaptation can exhibit both a stabilizing and a destabilizing effect on the overall closed loop system.
Future research directions will investigate the interaction between global homeostasis and distributed adaptation on nonlinear dendritic trafficking models with saturated compartments. We will also study how the network topology constrains the system behavior, as well as heterogeneous local set points.

\section{proofs}
\label{prf}

\emph{Theorem \ref{thm1}}: the proof is very similar to \textbf{Part 2} below. 
There is just a minor difference in the use of the matrix $\Omega_m$, which 
contains fixed elements $\omega_i^m$.

\emph{Theorem  \ref{thm:glocal}}: we split the proof in two parts, for readability. 
Part 1 and Part 2 below satisfy the conditions of 
\cite[Theorem 11.4]{khalil2002nonlinear}, which allows to conclude the exponential stability 
of the equilibrium of the closed-loop system from the analysis of two reduced subsystems.

 \textbf{Part 1: exponential stability of the boundary-layer system}

Define the function $\phi(g_i,\omega^{m}_i) = k_L(g_i-\bar{g}_i) + \phi_{\ell b}(\omega_i,\bar\omega)$.
The boundary layer system is obtained by finding the (parameter dependent) equilibrium $\bar\omega^{m}_{i}(g_i)$
of \eqref{w_compact}, which corresponds to the solution of the equation
 $\omega^{m}_{i} = \frac{\phi(g_i,\omega^m_i)}{\gamma_{L}}$ and belongs to the domain
 $\bar\omega^{m}_{i}(g_i) \geq \bar\omega+ \varepsilon_{\bar\omega}$.

Consider the new coordinate  $z_{i} = \omega^{m}_{i} - \frac{\phi(g_i,\omega_i)}{\gamma_{L}}$ for all $i \in \mathbb{I}$. Then,
\begin{align}
\dot{z_{i}}  
=& \dot{\omega}^{m}_{i} - \frac{d}{dt} \frac{\phi(g_i,\omega^{m}_i)}{\gamma_{L}} \nonumber \\
=&  -\frac{\gamma_{L}}{\epsilon}z_i 
-\frac{1}{\gamma_L} \left(
\frac{\partial \phi(g_i,\omega_i)}{\partial \omega_i}   \dot{\omega}_i + \frac{\partial \phi(g_i,\omega_i)}{\partial g_i}   \dot{g}_i 
\right)  \nonumber \\
=&  -\frac{1}{\epsilon}
 \left(\gamma_{L} - \frac{\partial \phi(g_i,\omega_i)}{\partial \omega_i} \right) z_i
 -\frac{1}{\gamma_L} \frac{\partial \phi(g_i,\omega_i)}{\partial g_i}   \dot{g}_i .
\end{align}

By introducing $\tau = \frac{t}{\epsilon}$, in the limit of $\epsilon = 0$, 
we obtain the boundary layer system 
$$
\frac{d z_{i} }{d\tau} = -\left(\gamma_{L} - \frac{\partial \phi(g_i,\omega_i)}{\partial \omega_i} \right)z_{i} \ , \quad \mbox{ for } \quad i \in \mathbb{I} \ .
$$ which is an exponentially stable system in the neighborhood of the equilibrium $\bar\omega^{m}_{i}(g_i) \geq \bar{\omega} + \epsilon_{\bar\omega}$
for $\gamma_{L} > \frac{\partial \phi(g_i,\omega_i)}{\partial \omega_i} = 0$.

\textbf{Part 2: exponential stability of the reduced-order system}

The reduced order system is obtained from \eqref{EQ:m_matrix}-\eqref{EQ:leaky-integrator}
by replacing the diagonal matrix $\Omega_m$ with the diagonal matrix $\bar{\Omega}_{m}(g)$
whose elements on the diagonal are given by $\bar\omega^{m}_{i}(g_i) \geq \bar{\omega}$,
whenever $i\in \mathbb{I}$, and by $\omega^{m}_{i,const}$, otherwise. 
We use a contraction argument to prove exponential stability of the reduced system.

\emph{Part 2.a: widened reduced system and linearization.}
First, consider the ``widened'' reduced order system obtained by 
replacing $\bar{\Omega}_{m}(g)$ with $\bar{\Omega}_{m}(q)$  where $q(\cdot):\mathbb{R} \to \mathbb{R}^n$
is any differentiable signal. The set of trajectories of the widened reduced system contains 
the original reduced system trajectories, corresponding to  additional constrain $q = g$.
Then, consider the linearized dynamics of the widened reduced order system
\begin{align}
\delta \dot{m} & = (L -\bar{\Omega}_{m}(q(t)))\delta {m}  + B\delta u \nonumber\\
\delta \dot{g}  & =  S\delta {m} - \Omega_{g}\delta {g}   \nonumber \\
\delta \dot{u}  & = -k_{G} \partial\psi(c^{T} g)c^{T}\delta {g} - \gamma_{G} \delta u \nonumber
\end{align}
where
$\psi$ corresponds to the function arising from the composition of \eqref{eq:G}, \eqref{EQ:V} and \eqref{EQ:Ca}. In what follows we construct a differential Lyapunov function to show exponential contraction of the system (\cite{forni2013differential}), which implies exponential stability. 

\emph{Part 2.b: diagonal Lyapunov matrix $P$ for the transport sub-dynamics.}
Consider the system
$\dot{\eta} = (L - \rho I) \eta$ where $\rho>0$ is a generic real constant.
Following Chapter 4 in \citep{Farina2000}, this system is positive
(off-diagonal elements are non-negative) and irreducible (by the connectedness
assumption on the neuron topology and the fact that $l_{ij} \neq 0$ iff $l_{ji} \neq 0$, for all $i \neq j$). 
Furthermore, each column of $L$ sums to zero
therefore each column of $L - \rho I$ sums to $-\rho$. As a consequence, the system
has a dominant eigenvalue in $-\rho$. All other eigenvalues have smaller negative real
part. Thus, by Theorem 15 in \citep{Farina2000}, for any $\rho>0$ 
there exists a positive definite and diagonal matrix $P$ such that 
$(L - \rho I)^TP + P (L - \rho I) < 0$. Take $P$ to be any positive diagonal solution
to this inequality for $\rho =  \frac{1}{2} \bar{\omega}$,
which ensures $\rho \leq \frac{1}{2}\lambda_{\min} (\bar{\Omega}_{m}(q(t)))$, for all $t$.

\emph{Part 2.c: Contraction and exponential stability }
Consider now the differential Lyapunov function 
$$V = \frac{\rho_{m}}{2}\delta m^{T}P\delta m + \frac{\rho_{g}}{2}\delta g^{T}\delta g  + \frac{1}{2}\delta u^{T} \delta u .$$
We have
\begin{align}
\dot{V}  = \,
& 1/2 \rho_{m}\delta m ^{T}(L^{T}P + PL)\delta m - \rho_{m}\delta m^{T}P\bar{\Omega}_{m}(q(t))\delta m  \nonumber \\
& + \rho_{m}\delta m^{T}P\delta u + \rho_{g}\delta m^{T}S\delta g - \rho_{g}\delta g^{T} \Omega_{g}\delta g \nonumber \\
&    -k_{G}\partial \psi (c^{T} g)c^{T}\delta {g}\delta u -\gamma_{G}\delta u^{T}\delta u \nonumber \\
= \,&  1/2 \rho_{m}\delta m ^{T}( (L - \rho I)^TP + P(L - \rho I))\delta m  \nonumber  \\
& - \rho_{m}\delta m^{T}P(\bar{\Omega}_{m}(q(t))-\rho I)\delta m \nonumber \\
& + \rho_{m}\delta m^{T}P\delta u + \rho_{g}\delta m^{T}S\delta g - \rho_{g}\delta g^{T} \Omega_{g}\delta g \nonumber \\
&    -k_{G}\partial \psi (c^{T} g)c^{T}\delta {g}\delta u -\gamma_{G}\delta u^{T}\delta u \ . \nonumber 
\end{align}
Then, using $(L - \rho I)^TP + P(L - \rho I)<0$, can be 
bounded by the following inequality
\begin{align}
\dot{V}
\le  & - \begin{bmatrix}
|\delta m| \\
|\delta g| \\
|\delta u|
\end{bmatrix}^{T}\underbrace{\begin{bmatrix}
	\begin{matrix}
	\lambda_{1}
	\end{matrix}
	& \rvline & -\lambda_{2}  & \rvline & -\lambda_{6}\\
	\hline
	-\lambda_{2} & \rvline &
	\begin{matrix}
	\lambda_{3}
	\end{matrix}
	& \rvline & -\lambda_{4} \\
	\hline
	 -\lambda_{6} & \rvline &
	-\lambda_{4} & \rvline & \lambda_{5}
	\end{bmatrix}}_{Q} \begin{bmatrix}
|\delta m| \\
|\delta g| \\
|\delta u|
\end{bmatrix}   \nonumber
\end{align}
where 
$
\lambda_{1}  = \frac{\rho_m}{2} \bar{\omega} \leq \frac{\rho_m}{2}\lambda_{\min} (\bar{\Omega}_{m}(q(t)))$ for all $t$, 
$\lambda_{2} = \frac{\rho_{g}}{2} |S| $,
$\lambda_{3}  =\rho_{g}\lambda_{\min}(\Omega_{g}) $,
$\lambda_{4}  = \frac{k_{G}}{2} |\partial \psi(c^{T}  g)c^{T}| $,
$\lambda_{5}  = \gamma_{G} $,
$\lambda_{6} =   \frac{\rho_{m}}{2} |P| $.

The problem now reduces to proving that $Q>0$. By Sylvester's criterion, the above matrix is positive-definite provided that its leading principal minors are positive. 
For the first minor is 
\begin{equation}
\label{eq:cond1}
\frac{\rho_m}{2} \bar{\omega} > 0 \iff \rho_{m} > 0
\end{equation}
For the second minor we get
\begin{align}
\label{eq:cond2}
\frac{1}{2}\rho_{m}\rho_{g}\lambda_{\min}(\Omega_{g}) \bar{\omega}  - \frac{\rho_{g}^2|S|^2}{4}  > 0 \nonumber \\
\iff {\rho_{m}  > \frac{|S|^2}{2\lambda_{\min}(\Omega_{g})\bar{\omega} }\rho_{g}}
\end{align}
For the third minor we get
\begin{align}
\frac{\rho_m}{2}\bar{\omega}  \left(\gamma_{G}\lambda_{\min}(\Omega_{g})\rho_{g} - \frac{k^2_{G}}{4}|\partial \psi(c^{T}  g)c^{T}|^{2}\right) + \nonumber\\
+ \frac{\rho_{g}}{2}|S|\left(-\frac{\gamma_{G}}{2}|S|\rho_{g} - \frac{1}{4}k_{G}|\partial \psi(c^{T}  g)c^{T}| \rho_m|P|\right) + \nonumber\\
-\frac{\rho_{m}}{2}|P|\left(\frac{1}{4}k_{G}|\partial \psi(c^{T}  g)c^{T}| |S|\rho_{g} + \frac{1}{2}\lambda_{\min}(\Omega_{g})|P|\rho_{g}\rho_{m}\right) > 0 , \nonumber
\end{align}

which  can written as
\begin{subequations}
\begin{align}
& \lambda_{\min}(\Omega_{g})\left(\frac{\gamma_{G}}{4} \bar{\omega} \rho_{m}\rho_{g} 
	- \frac{1}{4}|P|^{2}\rho_{g}\rho_{m}^2\right) \label{eq:bi1}\\
	+ &
\left(\frac{\gamma_{G}}{4} \bar{\omega} \lambda_{\min}(\Omega_{g})\rho_{m}\rho_{g} 
	- \frac{1}{4}\gamma_{G}|S|^2\rho_{g}^2\right)  \label{eq:bi2}\\
	> &
\frac{1}{4} \left( k_{G} |\partial \psi(c^{T} \!\! g)c^{T}||S||P|\rho_{g}\rho_{m}\right) \label{eq:bi3}\\
+ &
\frac{1}{4} \left( \frac{k_{G}^2}{2} |\partial \psi(c^{T}  g)c^{T}|^2\rho_{m}\bar{\omega} \right)
\label{eq:bi4}
\end{align}
\end{subequations}
In order for the above inequality to hold, we need sum of the terms \eqref{eq:bi1} and \eqref{eq:bi2} 
to dominate that of \eqref{eq:bi3} and \eqref{eq:bi4} . As a first step,
we observe that  \eqref{eq:bi3} and \eqref{eq:bi4} can be made arbitrarily small by setting $0<k_{G}\ll1$. Thus, what remains to show 
is that the terms of \eqref{eq:bi1} and \eqref{eq:bi2}  are positive; this can be guaranteed by satisfying the following relations:
\begin{align}
\label{eq:cond3}
\text{from \eqref{eq:bi1}  :} & \quad {\rho_{m} < \frac{\gamma_{G} \bar{\omega}}{|P|^{2}}}  \\
\label{eq:cond4}
\text{from \eqref{eq:bi2}:} &\quad  {\rho_{m} > \frac{ |S|^2}{\bar{\omega} \lambda_{\min}(\Omega_{g})}\rho_{g}} \ .
\end{align}
Therefore, combining \eqref{eq:cond1}, \eqref{eq:cond2}, \eqref{eq:cond3}, and \eqref{eq:cond4},
we conclude that the widened reduced system satisfies 
$\dot{V} \leq -\bar{\lambda} V$ for some $\bar{\lambda}>0$, whenever $k_{G}$ is sufficiently small
and $\rho_{m},\rho_{g},k_{G} > 0$ satisfy
\begin{equation}
	\dfrac{\gamma_{G}\bar{\omega}}{ |P|^{2}} > \rho_{m}  >\dfrac{ |S|^2}{\bar{\omega} \lambda_{\min}(\Omega_{g})}\rho_{g}.
\end{equation}

The exponential decay of the differential Lyapunov function guarantees
incremental exponential stability of the widened reduced system \citep[Theorem 1]{forni2013differential}.
This implies exponential stability of the reduced system.

\bibliography{ifacconf}

\begin{thebibliography}{16}
\providecommand{\natexlab}[1]{#1}
\providecommand{\url}[1]{\texttt{#1}}
\providecommand{\urlprefix}{URL }
\expandafter\ifx\csname urlstyle\endcsname\relax
  \providecommand{\doi}[1]{doi:\discretionary{}{}{}#1}\else
  \providecommand{\doi}{doi:\discretionary{}{}{}\begingroup
  \urlstyle{rm}\Url}\fi

\bibitem[{Aljaberi et~al.(2019)Aljaberi, O'Leary, and
  Forni}]{aljaberi2019qualitative}
Aljaberi, S., O'Leary, T., and Forni, F. (2019).
\newblock Qualitative behavior and robustness of dendritic trafficking.
\newblock \emph{IEEE 58th Annual Conference on Decision and Control (CDC)}.

\bibitem[{Bressloff(2009)}]{bressloff2009cable}
Bressloff, P.C. (2009).
\newblock Cable theory of protein receptor trafficking in a dendritic tree.
\newblock \emph{Physical Review E}, 79(4), 041904.

\bibitem[{Burute and Kapitein(2019)}]{Burute_2019}
Burute, M. and Kapitein, L.C. (2019).
\newblock Cellular logistics: Unraveling the interplay between microtubule
  organization and intracellular transport.
\newblock \emph{Annual Review of Cell and Developmental Biology}, 35(1),
  29--54.
\newblock \doi{10.1146/annurev-cellbio-100818-125149}.

\bibitem[{Costa-Mattioli et~al.(2009)Costa-Mattioli, Sossin, Klann, and
  Sonenberg}]{costa2009translational}
Costa-Mattioli, M., Sossin, W.S., Klann, E., and Sonenberg, N. (2009).
\newblock Translational control of long-lasting synaptic plasticity and memory.
\newblock \emph{Neuron}, 61(1), 10--26.

\bibitem[{Farina and Rinaldi(2000)}]{Farina2000}
Farina, L. and Rinaldi, S. (2000).
\newblock \emph{Positive linear systems: theory and applications}.
\newblock Pure and applied mathematics (John Wiley \& Sons). Wiley.

\bibitem[{Fernandez-Moya et~al.(2014)Fernandez-Moya, Bauer, and
  Kiebler}]{fernandez2014meet}
Fernandez-Moya, S.M., Bauer, K.E., and Kiebler, M.A. (2014).
\newblock Meet the players: local translation at the synapse.
\newblock \emph{Frontiers in molecular neuroscience}, 7, 84.

\bibitem[{Fonkeu et~al.(2019)Fonkeu, Kraynyukova, Hafner, Kochen, Sartori,
  Schuman, and Tchumatchenko}]{fonkeu2019mrna}
Fonkeu, Y., Kraynyukova, N., Hafner, A.S., Kochen, L., Sartori, F., Schuman,
  E.M., and Tchumatchenko, T. (2019).
\newblock How m{RNA} localization and protein synthesis sites influence
  dendritic protein distribution and dynamics.
\newblock \emph{Neuron}, 103(6), 1109--1122.

\bibitem[{Forni and Sepulchre(2013)}]{forni2013differential}
Forni, F. and Sepulchre, R. (2013).
\newblock A differential {L}yapunov framework for contraction analysis.
\newblock \emph{IEEE Transactions on Automatic Control}, 59(3), 614--628.

\bibitem[{Glock et~al.(2017)Glock, Heum{\"u}ller, and Schuman}]{glock2017mrna}
Glock, C., Heum{\"u}ller, M., and Schuman, E.M. (2017).
\newblock mrna transport \& local translation in neurons.
\newblock \emph{Current opinion in neurobiology}, 45, 169--177.

\bibitem[{Khalil(2002)}]{khalil2002nonlinear}
Khalil, H.K. (2002).
\newblock Nonlinear systems.
\newblock \emph{Upper Saddle River}.

\bibitem[{Marder and Goaillard(2006)}]{marder2006variability}
Marder, E. and Goaillard, J.M. (2006).
\newblock Variability, compensation and homeostasis in neuron and network
  function.
\newblock \emph{Nature Reviews Neuroscience}, 7(7), 563--574.

\bibitem[{O'Leary et~al.(2010)O'Leary, van Rossum, and
  Wyllie}]{o2010homeostasis}
O'Leary, T., van Rossum, M.C., and Wyllie, D.J. (2010).
\newblock Homeostasis of intrinsic excitability in hippocampal neurones:
  dynamics and mechanism of the response to chronic depolarization.
\newblock \emph{The Journal of physiology}, 588(1), 157--170.

\bibitem[{O'{L}eary et~al.(2013)O'{L}eary, Williams, Caplan, and
  Marder}]{o2013correlations}
O'{L}eary, T., Williams, A.H., Caplan, J.S., and Marder, E. (2013).
\newblock Correlations in ion channel expression emerge from homeostatic tuning
  rules.
\newblock \emph{Proceedings of the National Academy of Sciences}, 110(28),
  E2645--E2654.

\bibitem[{O'Leary et~al.(2014)O'Leary, Williams, Franci, and
  Marder}]{o2014cell}
O'Leary, T., Williams, A.H., Franci, A., and Marder, E. (2014).
\newblock Cell types, network homeostasis, and pathological compensation from a
  biologically plausible ion channel expression model.
\newblock \emph{Neuron}, 82(4), 809--821.

\bibitem[{Turrigiano(2008)}]{turrigiano2008self}
Turrigiano, G.G. (2008).
\newblock The self-tuning neuron: synaptic scaling of excitatory synapses.
\newblock \emph{Cell}, 135(3), 422--435.

\bibitem[{Williams et~al.(2016)Williams, O'Donnell, Sejnowski, and
  O'Leary}]{williams2016dendritic}
Williams, A.H., O'Donnell, C., Sejnowski, T.J., and O'Leary, T. (2016).
\newblock Dendritic trafficking faces physiologically critical speed-precision
  tradeoffs.
\newblock \emph{Elife}, 5, e20556.

\end{thebibliography}
\end{document}